# GPS Spoofing Attacks on Phasor Measurement Units: Practical Feasibility and Countermeasures

Fakhri Saadedeen, *Student Member, IEEE*, and Anamitra Pal, *Senior Member, IEEE*
School of Electrical, Computer and Energy Engineering
Arizona State University
Tempe, AZ, USA
Email: fsaadede@asu.edu; Anamitra.Pal@asu.edu

*Abstract* — Prior research has demonstrated that global positioning system (GPS) spoofing attacks on phasor measurement units (PMUs) can cripple power system operation. This paper provides an experimental evidence of the feasibility of such an attack using commonly available digital radios known as software defined radio (SDR). It also introduces a novel countermeasure against such attacks using GPS signal redundancy and low power long range (LoRa) spread spectrum modulation technique. The proposed approach checks the integrity of the GPS signal at remote locations and compares the data with the PMU's current output. This countermeasure is a ready-to-deploy system that can provide an instant solution to the GPS spoofing detection problem for PMUs.

## I. Introduction

Phasor measurement units (PMUs) are widely used in the electric power grid for wide-area monitoring, protection, and control, as they provide utilities with the time-stamped knowledge of when and where an event occurs. PMUs provide this time-stamp using the signals they receive from the global positioning system (GPS) satellites. GPS based solutions offer a wide geographic coverage, accuracy, cost effectiveness, compactness, and low power operation [1]. GPS signals contain the time, date, and location information. The PMU combines this data with the voltage and current information at a given location in the grid to provide time-synchronized voltage and current phasors, frequency, and rate-of-change-of-frequency information.

The GPS employed by PMUs is the one intended for standard civilian use, which is not encrypted or encoded; hence, it is susceptible to spoofing attacks. A spoofing attack is the introduction of a non-authentic signal to misrepresent the data for nefarious purposes. To make PMUs robust to GPS-spoofing attacks, a ready-to-deploy solution is needed. The solution must be adaptable and easy to implement. This paper first performs an actual experiment to demonstrate how GPS spoofing attacks can be carried out. Subsequently, it introduces a novel countermeasure against GPS spoofing attacks including how it can be implemented in practice.

### A. Global Positioning System (GPS)

Each GPS satellite broadcasts two signals, a military signal, and a civilian signal. The latter is used by the vast majority of GPS users. The time information is embedded in each respective signal, as the Nav/System data provides the GPS receiver with information about the position of the satellites as well as precise timing data from the atomic clocks aboard the satellites. Each satellite has a unique identification code (C/A code) that is repeated every millionth of a second. The Nav/System information is combined with the C/A code and then modulated within a carrier wave. The received GPS could be described as:

$$r(t) = \sum_{k=1}^{n} H_k (2P_c)^{\frac{1}{2}} (C_k(t) D_k(t)) \cos(2\pi(f_{L1} + \Delta f_k)t) + n(t) \quad (1)$$

where $H_k$ and $P_c$ are the channel matrix for the number of satellite and the channel power, respectively. $C_k(t)$ stands for the spread spectrum sequence (C/A code) and $D_k(t)$ is the navigation data. $f_{L1}$ is the carrier frequency and delta $f_k$ is the doppler frequency shift. Figure 1 illustrates the L carrier wave frequency along with the mixed C/A code. Both are transmitted as L1 or L2 signals.

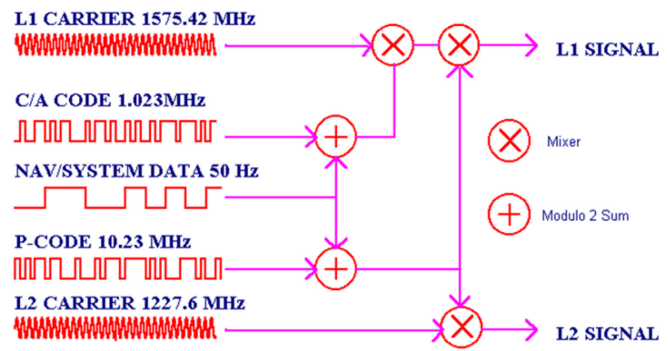

Figure 1. GPS signal (L1 and L2) illustration [2].

A non-encrypted GPS signal is susceptible to interferences and/or intentional spoofing attack. There are two types of spoofing attacks that can hinder PMU operation: (a) A signal interruption or interference attack that can be as simple as producing a jamming mechanism that prevents the GPS receiver of the PMU from collecting the signal from the satellite, or (b) A false data injection signal that may cause the PMU to send a faulty data to the system operator, such as a different phase angle or a different time stamp information. The latter has more serious consequences as it might trick the system operator into taking the wrong action. Spoofing attacks can have the following outcomes:

1. *They can reduce trust:* Any successful spoofing attack on PMUs will raise concerns regarding the trustworthiness of the synchrophasor infrastructure for operational decision-making.
2. *They can keep the user in the dark during an event:* For example, if an attacker plans to cause an event, he/she might cause a GPS interference, which keeps the origin of the event hidden from the system operator. Jamming the GPS signal may cause the onset time of the event to be unknown and a cascading event may follow the attack. Without knowing the time origin of the event, it can be very hard for power system engineers to get the complete picture of what happened during an event.

3. *They can provide incorrect phase angle:* Phase angle differences serve as effective indicators of system stress [3]. The effect of GPS spoofing attack on phase angle can be understood from the following equation,

$$P_{SR} \cong \frac{V_S \cdot V_R}{X_L} \cdot \sin \delta_{SR} \qquad (2)$$

where $P_{SR}$ is the active power flow between two buses across a transmission line, $V_S$ and $V_R$ are the sending-end and receiving-end bus voltage magnitudes, respectively, $X_L$ is the line impedance between buses, and $\delta_{SR}$ is the phase angle difference between bus voltage phasors at each line terminal. Equation (2) indicates that the angle difference relates voltage magnitudes, line impedance, and active power flow. A large angle difference can be a sign of topology change, heavy power flow between two buses, and/or big voltage drops. In [1], GPS spoofing resulted in a receiver clock offset error of 7:05ms time difference between the counterfeit and authentic GPS signals. This, in turn, caused the phasors to change ≈150°, which is big enough to cause faulty operation of the grid monitoring and control algorithms. Similarly, in [4], a GPS spoofing attack on the PMU was found to make the phase angle difference between two buses exceed 7°. This can activate load shedding algorithms, resulting in incorrect protection operation.

*B. Brief Summary of Prior Research*

The issue of GPS spoofing attacks on PMUs have been explored in prior research, but few demonstrate the actual process of a spoofing attack on a PMU without the need of a dedicated equipment and provide a complete solution besides simulated models.

Reference [1] investigated the effect of clock offset on PMU phase angle information but did not provide a solution to the problem. Instead, it referred to other publications that might have found a solution. Reference [5] analyzed the effect of a spoofing attack on PMUs and implemented a hardware as well as a software countermeasure. The hardware countermeasure consisted of deploying two antennas with different radiation pattern and comparing the standard deviation between the two. The paper showed an actual field test of this countermeasure. The software solution consisted of implementing a cross-layer detection algorithm that compares the time and phase angle between several PMUs in the network. The drawback of [5] is a lack of explanation regarding the location of the two antennas, i.e., it was not clear how the antennas themselves were made immune to the attack. Reference [6] proposed a software solution that consisted of measuring the phase angle and the current location and estimating them continuously. The drawback of this approach is that the spoofer can program the location to stay constant during the attack while a change in the phase angle can be construed as an actual change in the system. Reference [7] studied countermeasures to smart spoofers in which the authentic signal is received and then an altered version of this signal is sent to the target. The countermeasures consisted of a combination of multi-antenna spoofing discrimination and clock consistency check of GPS. However, [7] did not provide an actual hardware implementation of their solution.

In this paper, we present a novel hardware countermeasure to GPS timestamp alteration using a low-power wide area modulation technique. The proposed approach checks integrity of the GPS signal at remote locations and compares the data with a PMU's current output. This countermeasure is a ready-to-deploy system that can provide an instant solution to the GPS spoofing detection problem for PMUs.

## II. Spoofing Experiment Set-up

In the past, most spoofing experiments were conducted using a spoofer that was custom built for the sole purpose of spoofing a certain PMU or a GPS device. However, devices that allow wide band transmission of signals are commonly available and can be purchased online. To perform an actual attack and prove the feasibility of such an attack on PMUs, the easiest and simplest approach that an attacker might employ is to use a software defined radio (SDR). SDRs have the capability to receive and transmit data on a wide scale of frequencies. Most SDRs with transmission capabilities can produce a signal similar in length and waveform to an authentic GPS signal. The device chosen for this experiment was a HackRF SDR [8]. This SDR was chosen for its availability, capability, and reasonable price. Several parameters and data injection settings can be programed into the SDR for a spoofing attack, such as the location, time and date, and satellite positions.

The experiments were conducted on both PMUs and micro-PMUs. Figure 2 shows a micro-PMU's GPS data before spoofing was attempted. The settings show the actual number of satellites along with the latitude and longitude of the physical device location. As the location parameter is a unique indicator, a change in location would provide evidence of a successful spoofing attack. The HackRF SDR was able to conduct the attack, transmitting the minimum number of satellites to confirm a satellite lock by the micro-PMU. The micro-PMU was easily spoofed by the new signal and provided a GPS lock as well as showed the number of satellites it locked onto. Figure 3 shows the GPS data of the micro-PMU during the attack. The figure shows a location change proving that the attack is feasible. An actual attack will keep the PMU/micro-PMU's location the same and only change the time. This, in turn will change the phase angle as described in [4]. This experiment is equivalent to an actual attack in terms of the false time and date injection programed and produced by the SDR.

The test proved successful in spoofing a PMU using the brute force method in which the simulated signal overpowers the authentic signal and the receiver has no choice but to lock onto the spoofed signal as it becomes the only signal the receiver can detect. Another approach [9] is to start with a weak signal, then gradually increase the simulated signal, till it eventually takes over the authentic signal.

*A. Safety measures*

To safely conduct the type of experiment done above and to prevent the simulated signal from affecting other nearby equipment, a Faraday cage as shown in Figure 4 was built to accommodate both the simulated signal antenna and the antenna of the PMU/micro-PMU. The cage was layered with signal blocking materials that successfully blocked the simulated signal from escaping its boundaries. A cell phone

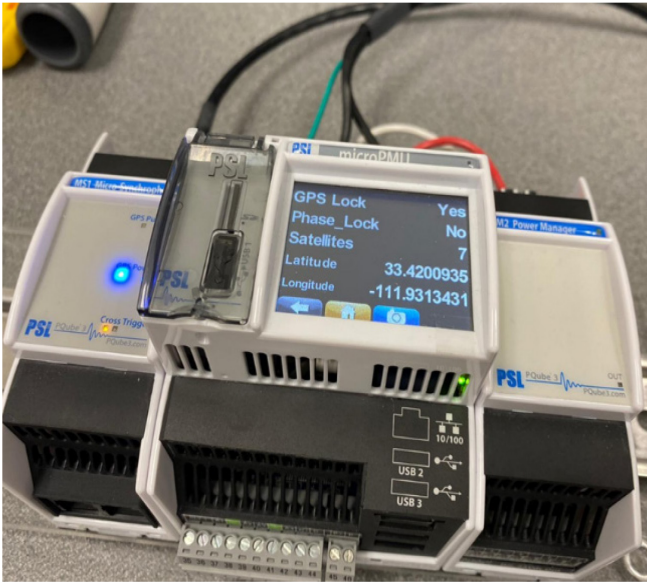
*Figure 2. micro-PMU GPS data before the attack.*

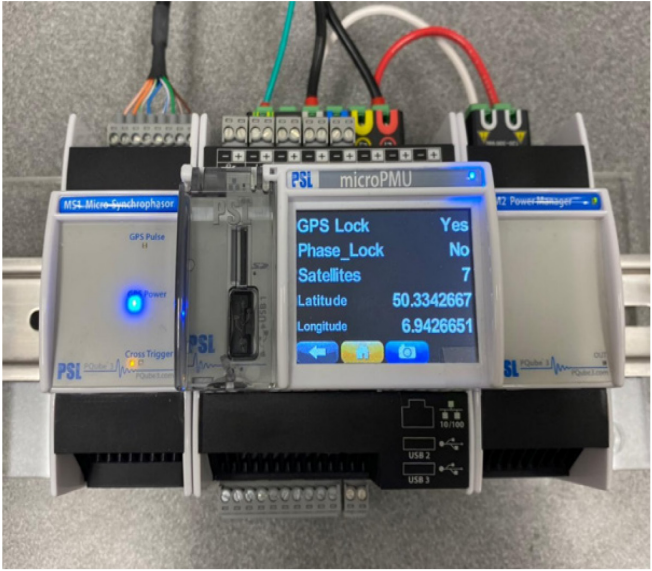
*Figure 3. micro-PMU GPS data during the attack.*

with GPS capability was placed right next to the box while the SDR was transmitting the spoofed signal. The phone did not detect the spoofed signal which shows the effectiveness of the Faraday cage. Along with the cage, a signal attenuator was placed in-line between the SDR and the sending antenna to further reduce the signal strength coming from the HackRF. The attenuator was rated at 10W 10dB which helped keep the signal contained inside the Faraday cage.

### III. Proposed Anti-spoofing System

We propose to build an anti-spoofing system using an encrypted low power long range wide area network protocol (LoRa) central receiving unit and multiple remote GPS receivers that are placed some distance away from the PMU. The GPS receivers will check the integrity of the signal at their current location and communicate it to each other and the central receiving unit through LoRa. The central unit will compare the GPS signal integrity received from the PMU's antenna vs. its own GPS antenna and determine if a spoofing attack has commenced or not. The remote GPS units can be configured differently depending on the need for distance from the central unit, power level needed to operate the devices, and the level of encryption implemented.

LoRa is a long-range low power spread spectrum modulation technique derived from chirp spread spectrum technology. It uses license-free sub-GHz radio frequency bands that enable long-range transmission with low power consumption. Thus, LoRa units are able to function well even when placed in remote locations.

The remote unit used here is the TTGO T-Beam LoRa module. It is low cost and has an 868/915 MHz LoRa transceiver and a neo-6m GPS receiver that is capable of receiving GPS signal as well as ephemeris data. It has wi-fi and Bluetooth capability for wireless communication. It is also equipped with a battery management system (BMS) for charging and discharging the 18650 battery [10] that can be added to the battery holder available with the unit. The onboard BMS makes it easy to add a miniature solar panel for powering the unit as well as using energy harvesting when the unit is located near power lines. The T-beam is an ESP32 system on a chip microcontroller that is programed through C. Note that more complex and robust LoRa solutions can be implemented for more stringent security needs.

#### A. Distance vs. Practicality

In this paper, the anti-spoofing remote units are placed 1-mile away from the PMU in three different directions. This ensures that even if one unit is affected by the spoofing attack, the other two will be able to indicate that an attack is in progress via majority vote (see Section IV). To validate the choice of a 1-mile radius, it is necessary to know the transmission distance of the SDR in relation to the distance of the anti-spoofer remote units from the central PMU location. As such, we determine the range up to which an SDR will be effective and show that a radius of 1-mile is sufficient.

Figure 5 shows where the anti-spoofing remote units might be placed in comparison to the PMU which is located inside a substation. The remote units are placed in different directions to lower the possibility of getting simultaneously impacted during a spoofing attack. PMUs use a satellite-synchronized clock that uses the GPS L1 band to receive the 1575.42 MHz signal. To find the wavelength of this signal, we use (3),

$$W = \frac{c}{f} \qquad (3)$$

where is $c$ is the speed of light and $f$ is the frequency. Hence, the wavelength of the L1 GPS signal is 0.190 m. The distance of transmission of any signal transmitter can be calculated based on the amount of power the device is using to transmit the signal along with the wavelength of the signal and antenna gain if applicable. The free-space path ratio is derived from the Friis transmission equation [11] which states that in a radio system consisting of a transmitting antenna transmitting radio waves to a receiving antenna, the ratio of radio wave power received, $P_r$, to the power transmitted, $P_t$ is,

$$\frac{P_r}{P_t} = D_t D_r \left(\frac{\lambda}{4\pi d}\right)^2 \qquad (4)$$

Therefore, we can calculate the power ratio decrease over distance (R) and compute a suitable distance between the PMU and the anti-spoofing remote units. Figure 6 shows the plot of the power ratio decrease in Watts (W) over the distance in meters (M) calculated using the Friis equation. It clearly demonstrates the value of a long-distance remote GPS antenna, as the signal of the spoofer degrades as the amount of power ratio decreases with (R). The calculation assumes a transmission power of 0.035W which is typical of most SDRs with transmission and receiving power gain of 10dBi. At this transmission power, the spoofer signal will be nonexistent at a 1-mile range, since the power of the spoofed signal is close to zero at that distance.

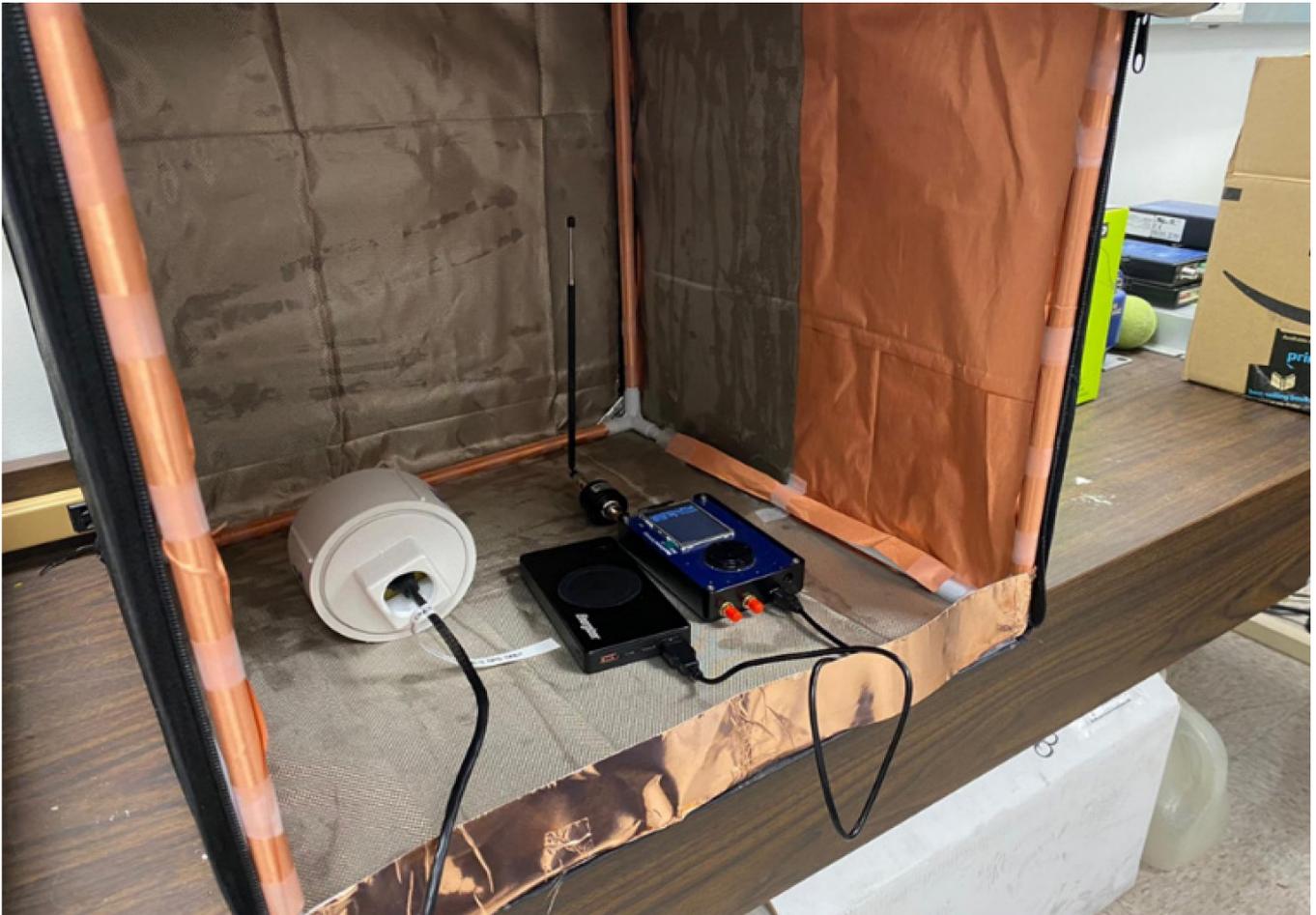

*Figure 4. Faraday cage.*

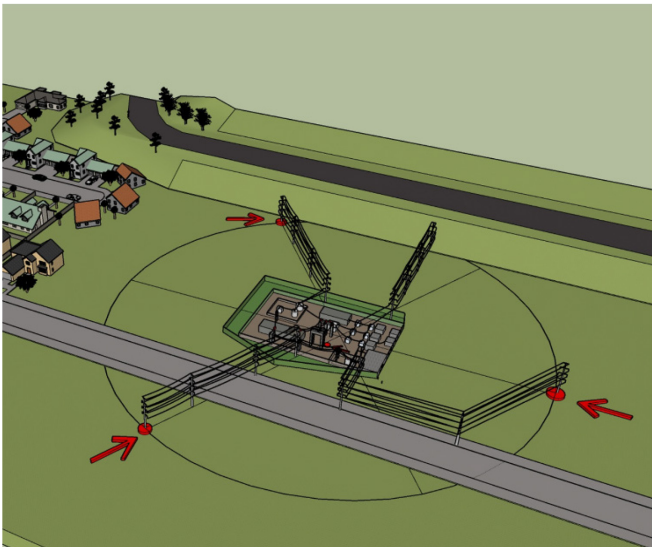

*Figure 5. Locations of remote units.*

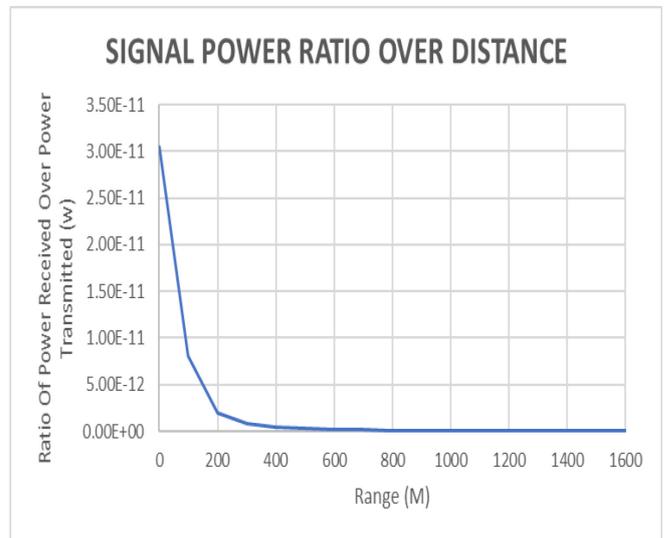

*Figure 6. Signal power ratio over distance.*

Note that higher the spoofer transmission power delivery, longer the distance the anti-spoofer remote units need to be, based on the free-space path loss equation. As such, the choice of the radius will be a tradeoff between how secure a utility needs its systems to be vs. the ease of installation and seamless integration of the remote units with the rest of the system. As such, the study conducted here provides a baseline for carrying out similar tradeoff studies by actual utilities.

## IV. RESULTS

Leveraging the LoRa module to develop a countermeasure for spoofing attacks is a novel development and advancement over previously developed methods that primarily used GPS signal authenticity through system redundancy [5]-[7], [12]-[15]. A signal redundancy check is a simple and effective way to counter a spoofing attack only if the anti-spoofer system itself is not impacted during the attack. Using LoRa adds robustness to the signal redundancy check as it introduces a signal on a different range modulation as LoRa uses 915 MHz while the L1 GPS uses 1575.42 MHz.

After programming certain parameters (e.g., bounded variations in the data of the anti-spoofing device), the proposed anti-spoofing system will be able to detect a spoofing attack when it occurs through a majority vote. This detection will allow the system operator to be aware of the spoofing attempt and discard the information presented from the attacked PMU. Alternately, the operator can use the anti-spoofing units to provide the accurate time stamp information during the spoofing attack.

Table I explains how an attack on a PMU will be detected by using the majority voting system of the remote anti-spoofing units. In the table, X indicates a scenario where a unit has not been impacted by a spoofing attack, while Y indicates a scenario where the spoofing attack has been successful in impacting the unit (e.g., by changing the unit's timing parameter information). Case 1 demonstrates normal operating conditions (no attack). In this case, the PMU has the same timing parameter information as the anti-spoofing units, and therefore the proposed anti-spoofing system states that no attack has been detected (last column of Table I). Case 2 demonstrates a detected spoofing attack as the PMU timing parameter is different from that of the anti-spoofing units. Case 3 demonstrates a spoofing attack in which the spoofer impacts the PMU as well as one of the anti-spoofing remote units (Remote Unit 1). However, this case will also be detected by the proposed system using the two remaining remote units. Note that the chances of more than one remote unit (along with the PMU) being simultaneously affected is minuscule due to the large distance between them. To summarize, the proposed anti-spoofing system will state that an attack has occurred if at least two remote units have different parameters than the PMU they are protecting.

**Table I:** *Detecting different cases of spoofing attacks*

| Case | PMU | Remote Unit 1 | Remote Unit 2 | Remote Unit 3 | Result |
|------|-----|---------------|---------------|---------------|--------|
| 1 | X | X | X | X | No Attack Detected |
| 2 | Y | X | X | X | Attack Detected |
| 3 | Y | Y | X | X | Attack Detected |

During the next set of experiments, one T-Beam remote GPS/LoRa unit was programed to receive L1 GPS signal through its onboard GPS receiver and send the gathered information to a selected LoRa receiving unit. Figure 7 shows the information that is being received by the T-Beam GPS module and constantly transferred to the receiving units. The information includes the latitude and the longitude of the device location, along with the number of visible satellites that the current unit sees, the visible satellites the device used to lock the signal, and finally the time transmitted from the satellite. More information can be obtained from the onboard GPS module such as date and the distance between the current device and the closest satellite, and can be programed to be received, displayed, and transmitted through the T-beam module. The number of packets that the sending unit is transmitting is also displayed for assurance and verification purposes. The current experiment involved fixing the location of the sending unit and varying the location of the receiving unit around the test area to evaluate the effectiveness of the LoRa signal. The signal strength can be measured by the receiving device while it is being moved from one location to another.

A second T-Beam was programed to receive the data that the first unit is sending to simulate the central receiving unit. The receiving unit will have the ability to receive multiple LoRa signals as a gateway interface. The second unit can display the received information as a verification process and also display the same information through a serial connection to a connected processing device. Figure 8 is showing the T-Beam receiving the information from the sending unit. The display shows the Received Signal Strength Indicator (RSSI) level, which estimates the power level being received from the sender. The higher the negative number displayed, the lower the power level received, which translates to a further distance measured from the transmitter. The unit is also displaying the size of each packet received. The unit can also display the number of packets being received as an additional option. Lastly, the unit is displaying the UTC time received from the sending unit which was received by the first unit's GPS module at a certain distance from the receiver.

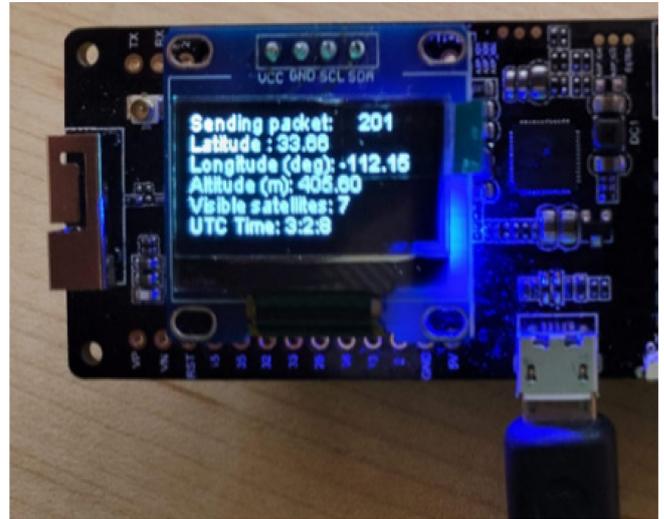

*Figure 7: Anti-spoofing remote unit displaying GPS information received through onboard GPS module.*

One important factor that greatly impacts signal strength is the antenna of the T-Beam LoRa transceiver. The antenna used in this experiment is an omni-directional antenna as it is intended to send and/or receive signals from all directions. While this antenna is suitable for this experiment, a dedicated unidirectional antenna will be more useful when the system is implemented in an actual setting. This is because a unidirectional antenna enhances the range of the signal transmitted since the power used to transmit will be concentrated in one direction (note that as the location of the remote units is fixed after installation, the direction of the transmitted signal will be known in advance).

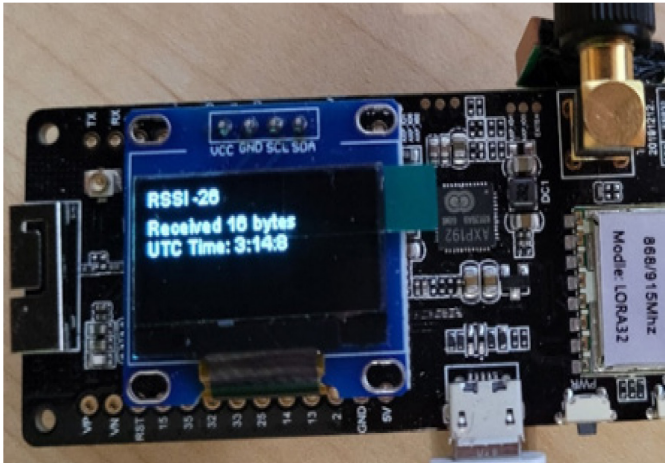

*Figure 8: Anti-spoofing receiver unit displaying UTC time information received through sender unit LoRa module.*

Figure 9 displays a snapshot of the received information available on a computing unit through the serial connection between the T-Beam's LoRa module and a personal computer (PC). The RSSI information displayed can be an indicator of the quality of the signal being sent on a constant basis as well as if the location of the remote unit is being compromised in any fashion. This information can be analyzed and compared with the time received from the PMU's GPS synchronization module that can be displayed through open-source software, such as OpenPDC. A simple comparison algorithm can find discrepancies between the two signals and sound the alarm if a spoofing attack is being attempted on the PMU in question.

## V. CONCLUSION

The problem of GPS spoofing on PMUs was considered in this paper and the dangers of such an attack discussed. An experiment was conducted to prove the feasibility of this problem using an SDR. A Faraday cage was built to contain the spoofed GPS signal and prevent it from impacting other devices during the testing. A novel countermeasure was introduced using GPS time redundancy and LoRa by authenticating the signal at a remote location and sending it to a central unit for comparison and verification.

Future work will include developing a software solution that will replace the compromised GPS signal measurement with the correct measurement by creating an interface between the PMU and the anti-spoofing system. A further study would include finding the minimum number of PMUs that need to be secured by the proposed anti-spoofing system to make all the PMUs of the network secure against GPS spoofing attacks.

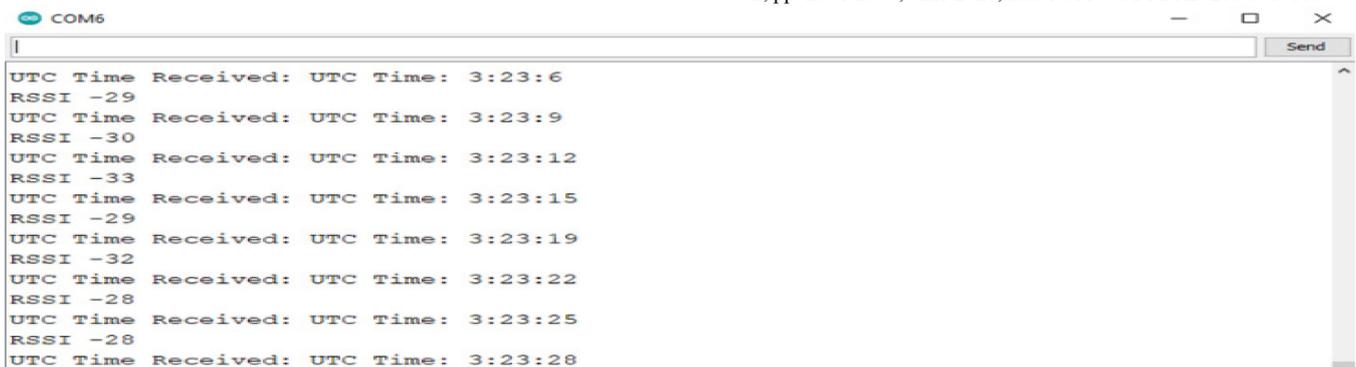

*Figure 9. Receiving unit displayed message through serial connection to a personal computer (PC).*